Special Topic：Electromagnetic metasurfaces: from concept to application

# Transformation optics based on metasurfaces

Chong Sheng[*], Hui Liu[†], Shining Zhu

National Laboratory of Solid State Microstructures and School of Physics, Collaborative Innovation Center of Advanced Microstructures, Nanjing University, Nanjing 210093, China

*e-mail: csheng@nju.edu.cn

†e-mail: liuhui@nju.edu.cn

The development of transformation optics and metamaterial provides a new paradigm for efficiently controlling electromagnetic wave. It is well known that a metamaterial is a novel artificial material with periodic structure, and the unit geometric size and distance between the units are much smaller than the working wavelength. Therefore, for an electromagnetic wave, a macroscopic system comprising these sub-wavelength units can be regarded as a novelty artificial material with a different permittivity, permeability, and refractive index compared to the original material. Furthermore, these electromagnetic parameters of the artificial material can be freely manipulated by modifying their unit structure, especially for a material that does not exist in nature. Among these materials, negative refractive index materials have been widely reported. For a negative refractive index material, when light is incident on its interface, the incident and refracted light are distributed on the same side of the surface normal, rather than on both sides as normal materials do. However, the field of transformation optics proposes the concept that the propagation of an electromagnetic wave in curved spacetime is identical in inhomogeneous and anisotropic optical mediums. Therefore, one can theoretically design the electromagnetic parameters of an artificial material to control the electromagnetic wave as well. The ongoing revolution of experimental technology, especially with regard to the advances in nanotechnology fabrication, has enabled the manifestation of imaginative transformation optics devices in experiments, such as invisibility cloaks in science fiction movies. However, most of transformation optics devices operate in microwave or near-infrared wavelengths, and there are still experimental challenges involved in working in the visible frequency. First, the artificial unit must be fabricated on a smaller scale using nanofabrication technology, which has challenges based on the current fabrication resolution. Material loss is another problem. The working unit of metamaterial is usually an artificial resonance structure of metal or photonic crystal. Metal structural units cannot avoid ohmic losses. For photonic crystal, it is difficult to overcome the scattering loss. These loss problems become increasingly serious, especially in the visible frequency.

Recently, new artificial material has been proposed to control an electromagnetic wave-metasurface, a two-dimensional metamaterial. Compared with a three-dimensional bulky metamaterial, this artificial plane material with sub-wavelength thickness greatly reduces fabrication time and mitigates fabrication complexity. Additionally, traditional metamaterials usually control the wavefront of an electromagnetic wave by accumulating the phase through propagating at a distance far larger than the wavelength. However, a metasurface can efficiently manipulate the wavefront of an incident electromagnetic wave through just the subwavelength propagation distance. Therefore, this can largely alleviate the propagation loss. Reference works about metasurfaces focusing on the far field manipulation of an electromagnetic wave have been extensively reported, such as planar optical devices based on metasurfaces and holographic imaging. In particular, one of the interesting works was done by Xiang Zhang's group [1], who produced a skin cloak in the visible frequency using a metasurface. Compared with traditional bulky cloaks, the skin cloak was only 80 nanometers thick. In addition, the reference works focusing on metasurfaces that control the near field regime of an electromagnetic wave have also been reported. For example, Lei Zhou's group [2] exploited a specific gradient-index metasurface to convert a freely propagating wave to a surface wave with nearly 100% efficiency (Fig. 1a). In Prof. Liu's work [3], by engineering the dispersion of surface plasmons with metasurfaces, the manipulation of surface plasmons with normal, non-divergent, or anomalous diffraction could be achieved (Fig. 1b). Related experimental work has been done by Alexander et al. [7] using a hyperbolic metasurface (Fig. 1c). Prof. Andrea Alù's group [8] used a graphene metasurface to produce topological transitions for the dispersion of a surface plasmon, whose contour was changed from a closed ellipse to an open hyperbola. Our group [6] experimentally demonstrated a one-dimensional curved metasurface to mimic bremsstrahlung radiation in curved space-time by controlling geometric phases of a surface plasmon (Fig. 1f). Additionally, a metasurface can also efficiently manipulate a waveguide mode. Prof. Chen [4] used a metasurface to demonstrate the asymmetric propagation of light without polarization limitations in a waveguide (Fig. 1d). In Prof. Yu's work [5], the researchers successfully exploited a metasurface to produce not only asymmetric optical power transmission but also waveguide mode converters and polarization rotators (Fig. 1e).

Given the fact that a metasurface has high manipulation efficiency for an electromagnetic wave in the near filed regime, our group [9] investigated experimental work on the analogy of gravity using a metasurface. According to transformation optics, the propagation of an electromagnetic wave in curved spacetime is identical in an inhomogeneous and anisotropic optical medium. The relationship between the metric of curved spacetime and the electromagnetic parameter is described as

$$\varepsilon = \mu = -\sqrt{-g}\, g^{ij}/g_{00}, \quad \omega_{0i} = g_{0i}/g_{00}, \tag{1}$$

where $g^{ij}$ is the metric of spacetime, $g = det(g_{ij})$, $\varepsilon$ is the electric permittivity, $\mu$ is the magnetic permeability, and $\omega_{0i}$ is the magneto-electric coupling coefficient. Given the equivalence, some general relativity phenomena in curved spacetime were emulated in the laboratory environment. We took one-dimensional topological defect-cosmic string as an example. As is well known, topology is a mathematical concept that has been extensively reported on in modern physics, ranging from condensed matter, high energy, and particle physics, to general relativity and modern cosmology. In particular, according to the cosmic inflation mode, the cosmic string, as a one-dimensional topological defect of spacetime, may have been generated in the early universe due to the spontaneously broken symmetry of the Higgs vacuum field predicted by theorists. Some type of astronomical observation, such as gravitational waves generated by the cosmic string, has been proposed to detect it. Although there has been a recent breakthrough in detecting gravitational wave generated by the merging of binary black holes, up to now there has been no direct astronomical observation evidence for the cosmic string.

To emulate one dimension topological defect-cosmic string (Fig. 2a), we first studied the spacetime metric of cosmic string, which is given as

$$ds^2 = dt^2 - dr^2 - \alpha^2 r^2 d\theta^2 - dz^2, \tag{2}$$

where $\alpha = 1 - 4Gm$, $G$ is the gravitational constant, $m$ is the mass density along the axis of the string, and we have used natural units. According to metric, the Riemann curvature can be calculated as $R_{12}^{12} = 2\pi(1 - \alpha)\delta^{(2)}(r)/\alpha$, where $\delta^{(2)}(r)$ is the two-dimensional delta function. Given the Riemann curvature, we know that cosmic string is flat everywhere except for a singularity in the center of the string. When the mass density parameter $m > 0$, the origin of the string has a positive curvature; however, if $m < 0$, the origin of the string has a negative curvature. Furthermore, according to the null geodesic line of a photon, the trajectory of a photon moves toward the string center with positive curvature, and it moves away from the origin of the string with negative curvature. What is more interesting is that the deflection angle $\Delta\theta = \pi(1-\alpha)/\alpha$. This scattering angle is dependent only on the mass parameter and independent of the impact parameter and momentum of the incident photon. The nontrivial scattering angle is drastically different from the scattering caused by the gravitational lensing of celestial bodies, which depends crucially on the parameter of an incident photon.

In order to emulate the cosmic string in a metasurface waveguide, according to Eq. (1), the effective electromagnetic parameter tensors in cylindrical coordinates are given as $\varepsilon_r = \mu_r = \alpha, \varepsilon_\varphi = \mu_\varphi = 1/\alpha, \varepsilon_z = \mu_z = \alpha$. In a metasurface waveguide, there are two orthogonal polarized waves: a transverse electric (TE) wave and a transverse magnetic (TM) wave. The TE wave has electromagnetic fields of ($E_\varphi$, $E_r$, $H_z$), so the refractive index is $\left(n_\varphi^2 = \epsilon_r \mu_z = \alpha^2, n_r^2 = \epsilon_\varphi \mu_z = 1\right)$. For the TM wave with the fields ($H_\varphi$, $H_r$, $E_z$), the refractive index is obtained as $\left(n_\varphi^2 = \mu_r \epsilon_z = \alpha^2, n_r^2 = \mu_\varphi \epsilon_z = 1\right)$. The refractive indexes for two polarizations have the same form, which is similar to that of a uniaxial crystal with a rotating axis along z-direction as $n = \begin{pmatrix} n_\varphi & 0 \\ 0 & n_r \end{pmatrix}$. As is well known, for an ordinary slab waveguide, the effective index is isotropic, and the iso-frequency is circular. However, if metasurface is used to weakly disturb the waveguide modes, the iso-frequency contours are changed from circles to ellipses, as shown in Fig. 2b. In order to achieve anisotropy in the rotating axis, the circular subwavelength grating is applied. The refractive index of such a metasurface waveguide can be given as $n^2 = R^{-1}(\varphi) \begin{bmatrix} n_e^2 & 0 \\ 0 & n_o^2 \end{bmatrix} R(\varphi)$, where $R(\varphi) = \begin{bmatrix} \cos(\varphi) & \sin(\varphi) \\ -\sin(\varphi) & \cos(\varphi) \end{bmatrix}$ is the rotation operator, and $n_o$ ($n_e$) is the effective index along the radial (azimuthal) direction that satisfies the iso-frequency contour: $(k_r/n_o)^2 + (k_\varphi/n_e)^2 = 1$. Furthermore, by comparing a refractive index required by the cosmic string, the ellipticity $\eta = n_e/n_o$ is taken as the same role as the mass density parameter $\alpha$ in the cosmic string. The TE mode with the ellipticity $\eta > 1$, as shown in Fig. 2b, corresponds to a negative topological defect at the origin, which repels light from the center of the string with a constant deflection angle. For the TM mode with ellipticity $\eta < 1$, the string has a positive mass density at the origin, and light is attracted to the origin of cosmic string, also with a constant deflection angle (Fig. 2a). In order to observe the definite photonic deflection in the nontrivial space of a cosmic string, the transverse electric (TE) mode was chosen in the experiment, and the results under three different impact parameters in the space of a negative topological defect with the mass density parameter $\eta \approx 1.065$ are shown in Fig. 2c, which had a constant deflection angle of about $\theta_1 = 11.0°$. Additionally, the green dashed lines in Fig. 2c are the geodesic lines of the cosmic string based on the theoretical calculation. All these results fit well. In addition, when the exciting transverse magnetic (TM) mode corresponded to a positive topological defect with a mass density parameter $\eta \approx 0.92$, a photon was attracted toward to the cosmic string with a constant angle of $\theta'_1 = -15.7°$ as shown in Fig. 2d.

Furthermore, by considering the photonic mode in this type of waveguide under the loss condition, we could emulate a symmetry breaking phase transition of the Higgs vacuum field during the inflation of the early universe, which generated cosmic string according to quantum mechanics. According to the above discussion, the non-degenerate TE and TM modes in the metasurface waveguide corresponded to different nontrivial spaces, and the two modes had a momentum mismatch $\Delta k = |k_{TE} - k_{TM}|$, where $k_{TE}$ ($k_{TM}$) is the momentum of the TE (TM) mode. Then we could define a coherent length $l_c = 1/\Delta k$. When considering the loss in PMMA, there was a propagation distance of $l_d = \max(l_{TE}, l_{TM})$, where $l_{TE}$ and $l_{TM}$ are the propagation distances for the TE and TM modes under the loss condition. According to Heisenberg's uncertainty principle $\Delta x \cdot \Delta p \geq \hbar/2$, if we could distinguish two modes, the critical momentum was about $\Delta p_{\min} \sim \hbar/l_d$. If $l_d > l_c$, then the momentum mismatch $\hbar \Delta k = \hbar/l_c > \Delta p_{\min}$, the two modes could be distinguished, and they could be assumed the symmetry breaking phase ($k_{TE} \neq k_{TM}$). However, if $l_d < l_c$, we had $\hbar \Delta k < \Delta p_{\min}$, and the TM and TE modes were non-distinguishable and considered one degenerate mode. This corresponded to the symmetry phase ($k_{TE} = k_{TM}$). When $l_c = l_d$, it could be considered the symmetry breaking phase transition point. Therefore, through tuning the loss in the metasurface waveguide, we could achieve a phase transition from trivial flat space to nontrivial topological space.

In summary, the metasurface, as a two-dimensional metamaterial, greatly reduced the propagation loss of electromagnetic waves and mitigated the requirements of fabrication compared with a three-dimensional bulky metamaterial. Therefore, it provides an effective tool for controlling electromagnetic waves. With the aid of transformation optics, a cosmic string as a one-dimensional topological defect generated during the early inflation of the universe, as well as the definite photonic deflection in the nontrivial space of a cosmic string, was successfully emulated in a metasurface waveguide. Furthermore, we can apply this novel topological property of definite deflection in imaging and energy transfer in further research. In addition, we emulated a quantum phase transition of the Higgs vacuum field using the symmetry breakage of a photonic mode under a loss condition in our optical system. We also believe that the combination of transformation optics and quantum optics in future research may provide a potential way to investigate the quantum nature of gravity.

Conflict of interest

The authors declare that they have no conflict of interest.

Acknowledgments

This work was financially supported by the National Natural Science Foundation of China (11690033, 61425018, 11621091 and 11704181), National Key R&D Program of China (2017YFA0303702) and National Key Research and Development Program of China (2017YFA0205700). C.S. gratefully acknowledges the support of the National Postdoctoral Program for Innovative Talents (BX201600070)

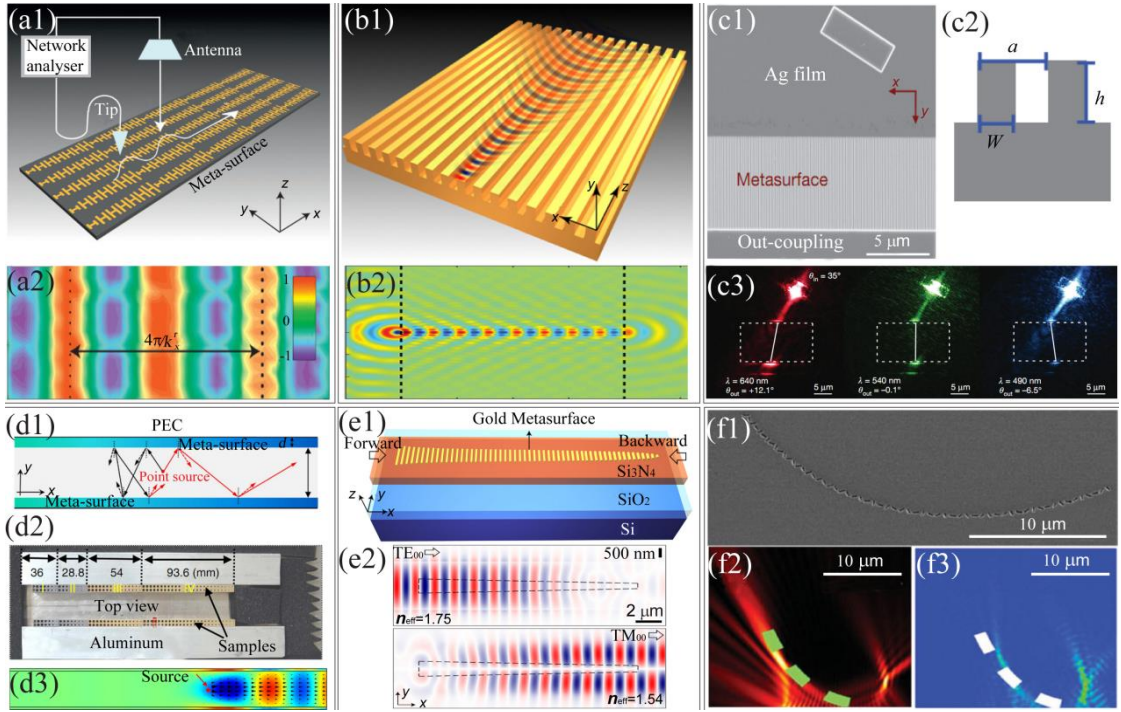

Fig. 1 Metasurface control of the electromagnetic wave in the near field regime. (a1)–(a2) A gradient-index metasurface used to convert a freely propagating wave to a surface wave. (a1) Schematic picture describing the near-field scanning technique; (a2) The experimental result using near-field scanning. Reproduced with permission from [2]. Copyright 2012 Springer Nature. (b1)–(b2) Metasurfaces for manipulating surface plasmons. (b1) Schematic illustration of a metasurface made of periodic metallic gratings; (b2) Propagation of surface plasmons along a metasurface exhibiting non-divergent diffraction. Reproduced with permission from [3]. Copyright AIP 2013 Publishing. (c1)–(c3) Measurement of the propagation of the SPP at a flat silver/hyperbolic metasurface interface. (c1) An SEM image of a device; (c2) Schematic of a hyperbolic metasurface; (c3) Images of the propagation of the SPP at the flat silver/ hyperbolic metasurface interface. The dashed boxes indicate the region of the hyperbolic metasurface. Reproduced with permission from [7]. Copyright 2015 Springer Nature. (d1)–(d3) Broadband asymmetric propagation of an electromagnetic wave without polarization limitations. (d1) Schematic diagram of a waveguide structure manipulated by a metasurface; (d2) The fabricated sample; (d3) The electromagnetic field distribution. Reproduced with permission from [4]. Copyright 2013 Springer Nature. (e1)–

(e2) The manipulation of waveguide modes using phase-gradient metasurfaces. (e1) Schematic of a working device; (e2) An electromagnetic field distribution demonstrating mode converts. Reproduced with permission from [5]. Copyright 2017 Springer Nature. (f1)–(f3) One-dimensional curved metasurface used to emulate bremsstrahlung radiation in curved space-time. (f1) SEM images of samples. (f2) SPP's rays via simulation; (f3) Experimental results. Reproduced with permission from [6]. Copyright 2018 American Physical Society.

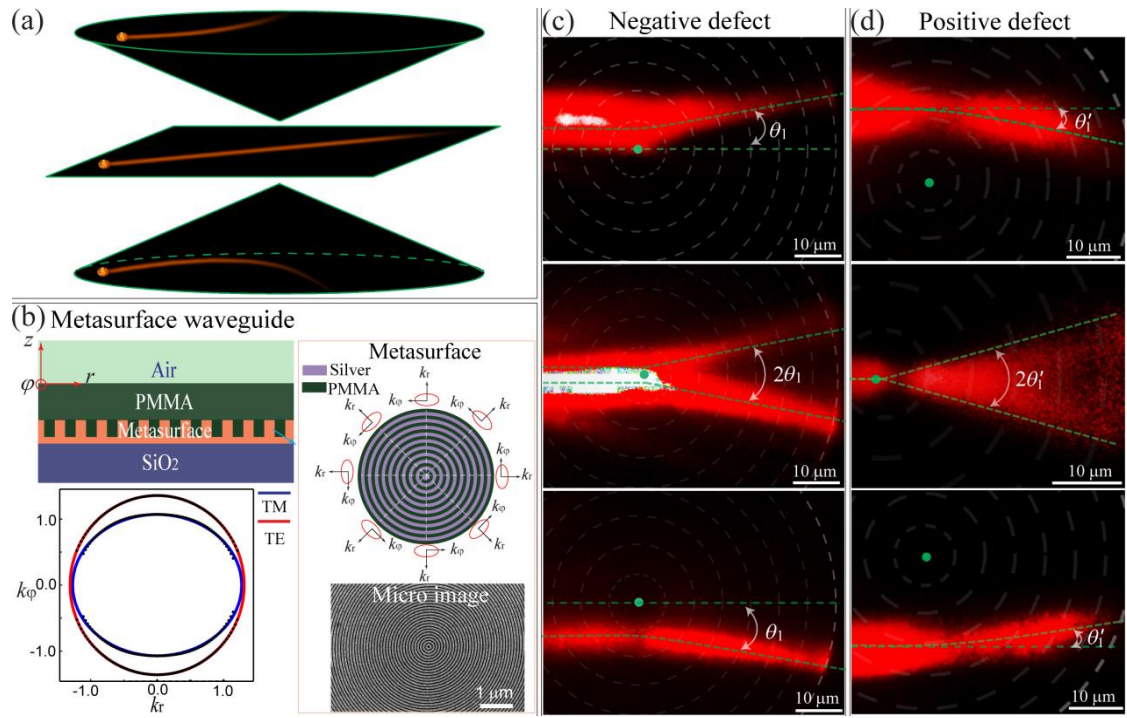

Fig. 2 Definite photon deflections in nontrivial topological spaces using metasurface waveguides. (a) When a photon propagates in the negative topological defect, it appears to undergo a repulsive force (top image); in the flat space, a photon propagates in a straight line (middle image); for the positive defect, it appears to attract the photon to the origin of the string (bottom image). (b) The schematic of a metasurface waveguide and the iso-frequency dispersion for TE and TM modes. For the TE mode, the ellipticity is $\eta > 1$, and it corresponds to a negative topological defect. For the TM mode, the ellipticity is $\eta < 1$, which represents a positive topological defect. (c) Experimental light deflection with a constant angle $\theta_1 = 11.0°$ in the negative topological defect with the mass density parameter $\eta \approx 1.065$ and different impact parameters ($b = 4.84, 0, -7.14$ μm). (d)

Experimental light deflection with a constant angle $\theta_1' = -15.7°$ in the positive topological defect with the mass density parameter $\eta \approx 0.92$ and different impact parameters ($b = 12.91, 0, -17.06$ μm). The green dashed line indicates the geodesic line based on the theoretical calculation, the series of dashed white circles represent the metasurface zones, and the green dot indicates the location of the topological defect.